\documentclass[10pt,a4paper,final]{iopart}

\usepackage{iopams}  
\usepackage{graphicx}
\usepackage{my_aas_macros}

\begin{document}

\title[Trends in Nuclear Astrophysics]{Trends in Nuclear Astrophysics}

\author{Hendrik Schatz$^{1,2}$}
\address{$^1$Department of Physics and Astronomy and National Superconducting Cyclotron Laboratory,
Michigan State University, East Lansing, MI 48824, USA}
\address{$^2$Joint Institute for Nuclear Astrophysics JINA-CEE, www.jinaweb.org}
\ead{schatz@nscl.msu.edu}

\begin{abstract}
Nuclear Astrophysics is a vibrant field at the intersection of nuclear physics and astrophysics that encompasses research in nuclear physics, astrophysics, astronomy, and computational science. This paper is not a review. It is intended to provide an incomplete personal perspective on current trends in nuclear astrophysics and the specific role of nuclear physics in this field. 
\end{abstract}

\vspace{2pc}
\submitto{\JPA}
%\maketitle

\section{Introduction}

Nuclear physics plays a special role in the cosmos. Everything that is visible in the night sky is powered by nuclear reactions. Nuclear physics governs the evolution of stars from birth to their final fate. And nuclear reactions that occurred in the past in the Big Bang, in stars and in stellar explosions have created every single chemical element found in nature today (with the exception of hydrogen). Astronomical observations offer also unique opportunities for nuclear physics investigations, by taking advantage of the extreme temperature and density conditions in stellar environments that are difficult or impossible to obtain in the laboratory. Examples include neutron star observations that provide the only avenue to investigate the properties of bulk, cold nuclear matter, or observations of neutrinos from supernovae that are messengers from the deep core of the explosions. Because of the central role that nuclear physics plays in the cosmos, nuclear astrophysics is not merely an activity that contributes nuclear physics data to astrophysics research, but is a field on its own at the forefront of science, with its own fundamental questions concerning the cosmic evolution of matter. The field encompasses research in nuclear physics, astrophysics, astronomy, and computational science. This paper is not a review. It is intended to provide an incomplete personal perspective on nuclear astrophysics and the specific role of nuclear physics in this field, with incomplete references to some examples and review papers. 

\section{Frontiers in Nuclear Astrophysics}

Nuclear astrophysics has come a long way since its beginnings in the early 20th century, largely coinciding with the birth of nuclear physics. Some of the prominent milestones of the field are the first proposals concerning nuclear energy generation in stars by Eddington in 1920 \cite{Eddington1920}, the identification of the nuclear reaction sequences powering the sun by von Weizs{\"a}cker in 1938 \cite{Weizsaecker1938}, Bethe and Critchfield in 1938 \cite{BeC1938}, and Bethe in 1939 \cite{Bethe1939},  the outline of some of the major element creating processes in stars by B$^2$FH \cite{B2FH} in 1956 and Cameron in 1957 \cite{Cameron1957}, and the detection of solar neutrinos in the late 1960s by Ray Davis \cite{Davis1968}. Despite of this long history there are a surprisingly large number of broad and fundamental open questions that the field currently seeks to address: of the 5 isotopes produced in the Big Bang, the discrepancy between the predicted $^7$Li production and primordial abundances inferred from observations remains unresolved \cite{Cyburt2015,Fields2014}. The composition of the Sun is a major open question: The amount of elements beyond He, the Sun's metallicity, had recently been revised  \cite{Asplund2009} from the canonical 2\% used as a reference standard for the field for decades down to 1.4\%. This is a remarkable change, that leads however to difficulties explaining helioseismology data. The origin of the portion of the heavy elements beyond about germanium that is not synthesized in the slow neutron capture process in red-giant stars (roughly half of the heavy elements), is not known \cite{Qian2012,Thielemann2011,Arnould2007}, despite of contrary claims in the non-scientific literature. For elements from tellurium to uranium the rapid neutron capture process (r-process) is responsible, but the site for this process is not known. Rapidly rotating magnetic supernovae \cite{Nishimura2015} or neutron star mergers \cite{Temis2015,Shen2015,Korobkin2012,Goriely2011} are some of the possible sites that are being discussed. The situation is even more unclear for elements in the germanium to palladium range \cite{Shibagaki2016,Montes2007}. It is likely that multiple processes contribute, including a weak s-process in massive stars, a weak r-process in neutrino driven winds either in neutron star mergers \cite{Martin2015} or supernovae \cite{Arcones2013}, a charged particle freezeout process \cite{Hallmann2013} or a so called neutrino-p process in core collapse supernovae \cite{Frohlich2006}. Another set of open questions concerns the basic mechanisms of stellar explosions. While core collapse supernovae are one of the major sources of lighter elements up to  iron, the mechanism that causes the transition of a core collapse into an explosion is not understood, nor is it known with certainty which stars explode as supernovae, and whether they leave neutron stars or black holes behind \cite{Janka2016}. Thermonuclear supernovae, the brightest stellar explosions, are thought to be explosions powered by the thermonuclear burning of an entire white dwarf star. However, the mechanism that would lead to an explosion of such a white dwarf star has not been identified \cite{Maoz2014}. Type I X-ray bursts, the most frequent stellar explosions observed, are thought to occur on the surface of neutron stars that accrete matter from a close-by companion star, but the response of these systems to changes in accretion rate and accreted composition is not understood \cite{Schatz2006a}. 

There are three main drivers that push the frontiers of the field and define the challenges for the future. (1) Multi-messenger observations are driving new questions and force the rethinking of old paradigms. Examples include dramatic progress in classic stellar composition observations through atomic spectral lines \cite{Frebel2015}, and in X-ray and radio observations of compact objects, most recently the detection of surprisingly massive neutron stars that rule out a large number of hypothesis regarding the nature of dense matter \cite{Demorest2010}. Supernova remnants are now routinely imaged across the entire electromagnetic spectrum, including MeV nuclear gamma ray lines. In addition, new types of messengers and observational approaches are beginning to be exploited. These include the detection of neutrinos from supernova 1987A and any future Galactic supernova \cite{Scholberg2012}, the advent of Asteroseismology \cite{Chaplin2013}, for example with the KEPLER mission, that enables to probe the structure of stellar interiors, and most recently the first direct detection of gravitational waves with LIGO \cite{Abbott2016}. (2) Multi-facility nuclear experiments and advanced nuclear theory efforts are providing critical nuclear physics data for stellar processes. However, despite decade long progress the vast majority of nuclear reaction rates in stars and in stellar explosions remain undetermined experimentally. A new generation of accelerator  facilities and techniques is poised to change this \cite{WhitePaper2016}. (3) Many of the processes in stellar interiors and in astrophysical explosions can only be understood through advanced computer models that follow fluid motions in all three spatial dimensions. As computational capabilities expand, more physically meaningful 3D simulations are now becoming possible for a wide range of astrophysical scenarios such as stellar surfaces including the sun \cite{Stein2012,Asplund2009}, advanced stellar burning stages such as oxygen burning zones \cite{Viallet2013}, hydrogen ingestion into stellar helium burning regions triggering neutron capture processes \cite{Woodward2015}, core collapse supernovae \cite{Janka2016}, and thermonuclear supernovae \cite{Ropke2011}. 

\begin{figure}[htb!]
 \centering
 \includegraphics[width=\textwidth]{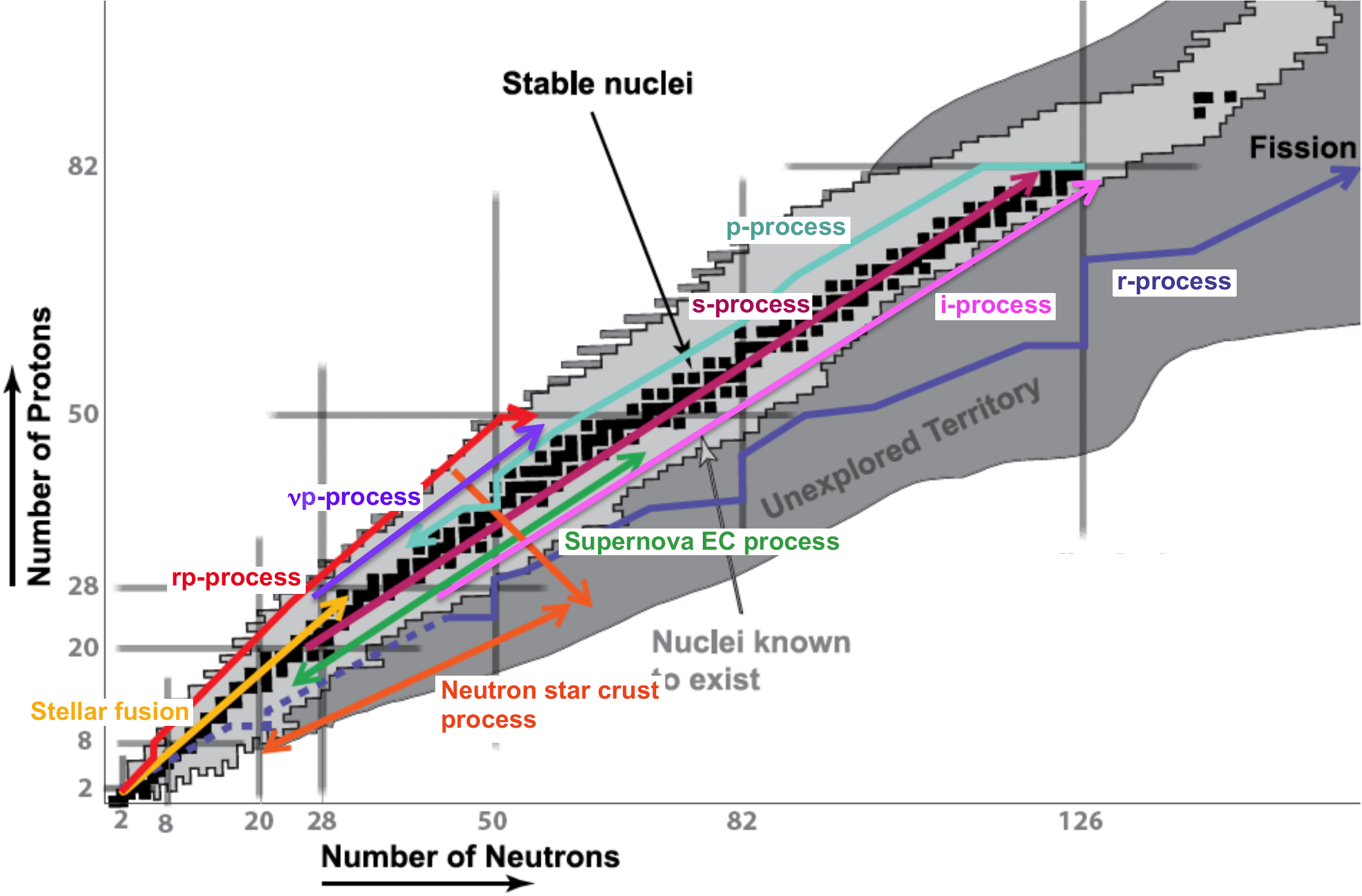}
 \caption{\label{FigChart}Schematic overview of the nuclear processes in the universe on the chart of nuclides (adapted from figure by F. Timmes).} 
\end{figure}

Taken together these technical developments create opportunities to address some of the long standing questions in nuclear astrophysics in the coming years. These questions, and the strategies to address them have been outlined recently in much more detail  in a white paper developed by the US nuclear astrophysics community in collaboration with their international colleagues \cite{WhitePaper2016}. In the following I briefly summarize some of the important trends in the field, with an emphasis on the importance of progress in nuclear physics. Accurate nuclear physics enables progress in nuclear astrophysics along three broad themes: it enables the validation of stellar models through comparison with observables, it enables to constrain stellar properties and system parameters, and it enables the prediction of nucleosynthesis contributions from individual processes and their historical evolution. The relevant nuclear processes are sketched in a schematic way in Fig.~\ref{FigChart}. 

\section{Nuclear Physics Enables Astrophysical Model Validation}

Progress in nuclear astrophysics requires the development of hypotheses and models that describe what happens deep inside stars and stellar explosions. These models must then be compared with data to determine whether they indeed reflect what happens in nature. Without such model validation, progress is not possible. The challenge is that data on stellar interiors cannot be obtained directly. Rather, in most cases only a limited amount of radiation from the stellar surfaces or the debris of the explosion can be measured (exceptions include the detection of neutrinos from nearby supernovae or new data on stellar oscillations that can be used for Asteroseismology \cite{Chaplin2013}). Nuclear physics opens a unique validation path for astrophysical models by linking such observables to the conditions in the deep interiors. This is done by exploiting the strong dependence of nuclear processes on temperature, density, and composition, but requires accurate nuclear data.  Once the nuclear physics is fixed, observables related to nuclear processes can be used to validate or falsify stellar models. 

In the future, model validation will become more important because of the increased complexity of stellar models, for example by moving from 1D to 3D calculations, or by including rotation. With more degrees of freedom in stellar models, validation will be key to ensure models reflect in a reasonable way what is actually happening inside a star or a stellar explosion. 

\subsection{Models for Neutron Capture Processes}

Validation through observed abundance patterns are of special importance for the neutron capture processes that produce elements beyond iron. These include the slow neutron capture process (s-process), the rapid neutron capture process (r-process), and possibly an intermediate neutron capture process (i-process). For these processes, detailed abundance patterns can be obtained from the solar system composition by de-convolution of multiple processes (s- and r-process), from abundance patterns in very metal poor stars that preserve signatures of neutron capture processes in the early Galaxy that polluted the material the star formed from (s, i, and r-process), and from meteoritic pre-solar grains (s-process, some r-process, novae, supernovae). These pre-solar grains were formed less than a billion years prior to solar system formation in stellar environments of the solar neighborhood, and found their way into the solar system, where they can now be analyzed in the laboratory \cite{Clayton2004}. 

{\em s-process:} The site of the s-process \cite{Kaeppeler2011}, which produces about half of the heavy elements beyond germanium by a sequence of slow neutron captures and much faster $\beta$-decays, is known since the first observation of technetium on the surface of a red giant star in 1954 \cite{Merrill1952}. Technetium is an element that does not have a stable isotope, and must therefore be created inside the star it is observed in. Based on nuclear physics arguments, the only plausible way to produce the neutrons for an s-process in such a stellar environment is via mixing between hydrogen burning and helium burning layers. Hydrogen burning layers contain protons and $^{14}$N from hydrogen burning via the CNO cycle. When mixed into the helium burning layer, which contains helium and the helium burning product $^{12}$C, proton capture on $^{12}$C and the subsequent $\beta$-decay of $^{13}$N produce $^{13}$C, while helium capture reactions and $\beta$-decays on $^{14}$N produce $^{22}$Ne. When immersed in helium rich environments, both isotopes act as strong neutron sources via ($\alpha$,n) reactions. S-process abundance observations are therefore intimately linked to mixing processes in stars, and can be used to directly probe quantitatively the mixing deep inside red giant stars. Uncertainties in the $^{13}$C($\alpha$,n) and $^{22}$Ne($\alpha$,n) reaction rates, as well as in the competing $\alpha$ capture rates, other potential neutron sources, and neutron absorbing reactions remain the main limitations of this approach and will have to be addressed in the future \cite{Vinyoles2016}. 

The s-process abundances itself also contain important information about the astrophysical environment. Neutron capture reactions in the main s-process are mostly in steady flow such that $Y <\sigma v>$ is constant (with $Y$ being the produced s-process abundance of an isotope along the reaction path, and $<\sigma v>$ is the stellar reaction rate). This means that the nuclear reaction rate $<\sigma v>$ has a strong influence on the abundances produced, but that  
the sensitivity on the astrophysical environment is limited as neutron capture rates depend only weakly on temperature, and because the reaction path is fixed to the valley of stability. However, the steady flow equilibrium breaks down when the reaction sequence splits into parallel branches. At these branch points neutron captures and $\beta$-decays occur with comparable rates. Isotopic abundance ratios along different branches therefore become very sensitive to  neutron density and, through the temperature sensitivity of the $\beta$-decay, on temperature. The experimental and theoretical determination of some of the key branch point rates enabled the validation of the today accepted multi-exposure s-process model in red giant (AGB) stars, where mixing processes activate both, the 
$^{13}$C($\alpha$,n) neutron source between thermal pulses, and the $^{22}$Ne($\alpha$,n) neutron source during helium flashes \cite{Kaeppeler2011}. With some experimental reaction rate data in place, branch point abundances could only be explained with both neutron exposures, and not with a single, average, neutron exposure, providing strong validation of the stellar models \cite{Kaeppeler1990}. 

The s-process branch points occur per definition at unstable nuclei. Measurements of neutron capture rates on these isotopes are difficult and many of the shorter-lived cases have not been determined experimentally. To maximize the validation potential of stellar models through s-process nucleosynthesis, measurements of neutron capture rates on these radioactive species are important for the future. In some cases, samples for such measurements could be provided by new radioactive beam facilities using isotope harvesting techniques, but depending on the half-lives this requires a nearby, or in house, neutron beam facility \cite{Reifarth2004}. 

{\em r-process:} The r-process is the second major nucleosynthesis process producing elements beyond germanium \cite{Qian2012,Thielemann2011,Arnould2007}. It proceeds via rapid neutron captures and $\beta$-decays, possibly including ($\gamma$,n) photodsintegration reactions and fission processes. Unlike for the s-process, the r-process site has not been determined with certainty. A large number of hypothetical models are therefore discussed in the literature. Among the more recently proposed sites are direct ejecta from neutron star mergers  \cite{Temis2015,Shen2015,Korobkin2012,Goriely2011}, 
neutrino driven winds in neutron star mergers \cite{Martin2015} or core collapse supernovae \cite{Arcones2013}, magneto-hydrodynamically driven jets in rapidly rotating supernovae \cite{Nishimura2015}, or a neutrino driven r-process in He shells of supernovae \cite{Banerjee2011}. In this situation, nuclear physics provides a possible pathway to validate models through comparison with observed r-process elemental abundances. The elemental abundance patterns produced by the r-process are well known from observations of metal-poor stars that preserve the composition of r-process events in the early Galaxy \cite{Frebel2015}. Observational constraints may in the future distinguish between supernovae and neutron star mergers, and first hints in favor of neutron star mergers are being obtained from the observations of optical afterglows of short $\gamma$-ray bursts, so called kilo-novae \cite{Tanvir2013}  and the r-process enrichments observed in some dwarf galaxies (see below) \cite{Ji2015}. However, nuclear physics seems to be the only way to validate individual sites and mechanisms within these broad scenarios.  
Unlike the s-process, the r-process reaction sequence includes very neutron-rich unstable isotopes. In low temperature scenarios, the so called cold r-process, neutron capture rates are comparable to $\beta$-decay rates along the r-process path. In the hot r-process, the path is determined by equilibrium between neutron capture and photo-disintegration. In either case, the path strongly depends on neutron density, in the hot case also on temperature, and is expected to change dramatically during the r-process. As the path sweeps over the chart of nuclides, characteristic abundance patterns are formed depending on the nuclear properties such as $\beta$-decay half-lives, $\beta$-delayed neutron emission probabilities, neutron capture rates (in the cold r-process case) and nuclear masses (in the hot r-process case) along a the momentary reaction sequence \cite{Kratz1993}. As the r-process reaches the actinide region and beyond, fission also plays a critical role \cite{Eichler2016}. Once the nuclear physics is known, this superposition of characteristic abundance patterns can be calculated for a given astrophysical model and then be compared with observations to distinguish between the proposed astrophysical environments and astrophysical assumptions. The goal is not to reproduce the observed abundances accurately (which requires to address nuclear and astrophysical uncertainties, with the latter often overwhelmingly large) but, by fixing the nuclear physics, to create validation pathways and guidance for the astrophysical aspect of the problem. A potential difficulty are freezeout effects. As the neutron density drops and the r-process ends some abundance features may be washed out, and new ones can be created for example through late time fission or preferential capture of neutrons in certain regions \cite{Eichler2016}. Constraints on the freezout behavior may also provide valuable validation opportunities, though. 

The long standing challenge for the field has been to produce the extremely unstable nuclei along the r-process path to enable measurements of the critical nuclear physics quantities. This is one of the primary motivations for a next generation of radioactive beam facilities such as RIBF, FAIR, and FRIB \cite{WhitePaper2016}. Despite of these difficulties, some experiments have succeeded to reach the r-process - examples include the half-life measurement of $^{130}$Cd at ISOLDE/CERN \cite{Kratz1986},  the half-life measurement of $^{78}$Ni at NSCL/MSU \cite{Hosmer2005}, mass measurements at ATLAS/ANL \cite{VanSchelt2013}, recent measurements near $N=126$ at FRS/GSI \cite{Caballero2015}, and the first results from a next generation radioactive beam facility at RIBF/RIKEN where a broad range of new half-lives in the r-process has been determined \cite{Lorusso2015}.  A similar challenge has been the theoretical prediction of nuclear properties along the r-process path with sufficient precision so as to be useful for astrophysical applications. R-process nuclei are heavy and in general, with a few exceptions near closed shells, well beyond the reach of large scale shell model calculations. While a number of very useful global models for the prediction of masses \cite{Lunney2003}, decay properties \cite{Marketin2015,Moller1997}, and reaction rates \cite{TALYS2008,Rauscher2000a},exist, their accuracy is not sufficient for astrophysical applications and their uncertainty is difficult to be characterized in the absence of nuclear data. Indeed, differences in predictions of the various models indicate much larger uncertainties than a comparison with data near stability would imply (see \cite{Lunney2003} for masses, or \cite{Beun2009} for neutron capture rates). There are however a number of very promising theoretical developments, including Density Functional Theory approaches, which also provide estimates for uncertainties. Examples include predicted masses \cite{Erler2012} or fission processes \cite{Sadhukhan2016}. To be useful for astrophysical applications it will be important to create global data sets with theoretical predictions for all nuclei (the entire mass and element range, and also odd-even and odd-odd nuclei), even if that means to compromise the quality of the prediction in some areas. This will be essential to ensure progress in nuclear theory translates into progress in nuclear astrophysics. 

{\em i-process:}  Recently, an intermediate neutron capture process (i-process) with neutron capture timescales inbetween the s- and the r-processes has been proposed to occur in red giant stars, especially in old stars with low metal content, when hydrogen and helium burning zones mix more easily, leading to entrainment of hydrogen into helium burning regions \cite{Dardelet2015}. This has been motivated by the identification of these entrainment processes in 3D fluid dynamics calculations \cite{Woodward2015}. Validation of such new and surprising results requires the use of i-process nuclear data to predict the characteristic abundance signatures in these new astrophysical models. These signatures have to then be matched to observed signatures that cannot be explained with other processes. Two of such (potential) signatures  have been identified:  (1) The peculiar abundances measured on the surface of Sakurai's object, a star at the very end of its red-giant phase that is thought to have been already on a path towards becoming a planetary nebula when one more thermal pulse ignited the star for one last time \cite{Herwig2011}. (2) The abundances detected in a class of metal poor stars with strong carbon enhancement that exhibit heavy neutron capture elements that neither fit a solar pattern, an r-process pattern, or a s-process pattern. The standard explanation has been to invoke a mix of s- and r-processes, but at least in some cases it has been difficult to obtain a good fit of the observations with this approach. A recently proposed alternative explanation is that some or all of these stars may have been polluted by an i-process in the early Galaxy \cite{Dardelet2015}. 

While still somewhat speculative, it is very important for nuclear physics to contribute reliable data for i-process models so that models can be validated (or falsified) in the context of the observational signatures. This is the only way to determine whether the i-process indeed is responsible for the observed abundance patterns or not. In terms of nuclear physics needs, the i-process is unique in that it requires knowledge of neutron capture rates on unstable neutron deficient nuclei a few mass units away from stability. Determining neutron capture rates on unstable nuclei has been a formidable challenge for experimental and theoretical nuclear physics, because both, neutron and target nucleus are unstable and traditional experiments with a stable or unstable beam impinging on a stable target are not viable. A variety of techniques have been developed, but all have strong limitations and significant systematic uncertainties. These include experiments that inform input to statistical reaction rate calculations, such as measurements of low lying dipole strength or the new beta-Oslo method \cite{Spyrou2014}, transfer reactions \cite{Kozub2012}, Coulomb breakup \cite{Gobel2016}, or other so called surrogate techniques \cite{Escher2012}. The i-process problem, and the fact that most of the important reactions involve nuclei that are not too exotic and can be produced in significant quantities at current facilities, provides a strong motivation to push the development of these techniques now.

\subsection{Neutrinos}
With the detection of solar neutrinos in the 1960s and neutrinos from supernova 1987A, a new era of neutrino astronomy has begun. 
As neutrinos are messengers that probe the deepest interiors of stars and supernova explosions, their detection opens up exciting opportunities to directly validate models of stellar environments not easily accessible by other observational probes. Nuclear physics is key in this validation process as the neutrinos are produced by nuclear processes - the pp-chains and the CNO cycle in the Sun, and electron capture reactions in the collapsing core of a supernova. The validation of the standard solar model through the detection of solar neutrinos is one of the prime examples that illustrates the importance of accurate nuclear physics for validating stellar models. It was only through decade long work in nuclear theory and experiment that the nuclear reactions in the core of the sun were well enough understood to be able to demonstrate that the detected neutrino flux was lower than expected, the so called solar neutrino problem \cite{Bahcall1976}. This ultimately led, together with other observations, to the conclusive discovery of neutrino oscillations \cite{Ahmad2001} and, once the properties of neutrinos were understood, to a convincing validation of our understanding of the conditions in the center of the Sun.

The detection of neutrinos from a Galactic supernova would open up similar opportunities for the validation of core collapse supernova models, especially the explosion mechanism. Unfortunately these events are rare (a few per century at most) and so far none has occurred since neutrino detectors became operational. Current neutrino detectors would detect 1000s of neutrinos from such an event \cite{Scholberg2012}. An important goal to enable this validation pathway is to determine with sufficient precision the nuclear reactions that drive the neutrino emission and shape the expected neutrino signal. The relevant reactions have been recently identified \cite{Sullivan2016} and experimental and theoretical work is ongoing to address this nuclear physics need. One experimental approach is the use of charge exchange reactions that can simulate weak interaction processes and can be used to extract the relevant transition rates \cite{Cole2012}. 

\subsection{P-process models}

The p-process is responsible for the origin of 35 stable isotopes that are too neutron deficient to be produced by neutron capture processes \cite{Rauscher2013,Arnould2003}. Yet, nature has found a path to create these isotopes, albeit with typically factors of 10-100 lower abundance than other more neutron rich isotopes of the same element.  A number of sites have been proposed, including the outer layers of various types of type Ia supernovae  (white dwarf mergers, accreting white dwarfs that ignite because their mass exceeds the Chandrasekhar mass limit, or white dwarf explosions triggered by the ignition of an accreted helium layer) or the shock front passing through the outer layers of a star undergoing a core collapse supernova explosion. The common feature of most current p-process scenarios is that the p-nuclei are produced by photo-disintegration of a distribution of stable seed nuclei. These seed nuclei were produced by previous stellar generations, by an s-process in the exploding star or, in the case of accreting scenarios, in the companion star from which matter is transferred from \cite{Travaglio2011}. 

A challenge for current attempts to validate p-process models is that neither very many observations are available, nor the nuclear physics is well understood. The observational challenge is that stellar spectroscopy is insensitive to the minute amounts of p-isotopes contributing the the abundance of a particular element. This has traditionally left the solar abundance distribution as the only reference, which may be a superposition of multiple p-processes and may therefore be only of limited use to guide the identification of the site. Meteoritic abundances may provide the only other isotopic information and are therefore very valuable \cite{Rauscher2013}. For example, the isotopic composition of some meteorites provides strong evidence for the presence of long lived, radioactive p-process isotopes in the early solar system such as $^{92}$Nb and $^{146}$Sm, which have since decayed. On the nuclear physics side, key reactions have been identified for a core collapse supernova scenario and involve a large number of stable and unstable nuclides. While impressive progress has been made in measuring p-process reactions on stable nuclei (for example \cite{Gyurky2016}), the reactions on unstable isotopes remain uncertain. 

Because of these challenges, and because the p-process is rather robust owing to the small amount of material that needs to be produced, validation of p-process models has so far not allowed researchers to discriminate among a large range of models. Rather, many p-process models can reproduce most p-process abundances within typically a factor of 3. An exception are the light p-nuclides $^{92,94}$Mo and $^{96,98}$Ru which have unusually large isotopic abundances and are underproduced by most p-process models. An interesting aspect of the p-process is that it can be used to not only validate the temperature conditions in the explosive nucleosynthesis site, but also the initial composition of the seed nuclei. Indeed, recent models of the p-process in type Ia supernova have attempted to address the underproduction of $^{92,94}$Mo and $^{96,98}$Ru by assuming an enhancement in s-process elements in the companions star that triggers the explosion \cite{Travaglio2011}. This has only been partially successful as even with favorable parameter choices the large abundance of $^{94}$Mo remains a puzzle. The alternative of a rapid proton capture process or a $\nu$p-process serving as additional source of the enhancement of the light p-nuclei is disfavored by the similarly large enhancement of radioactive $^{92}$Nb, which is shielded from these processes but can be readily produced in a traditional photodisintegration based p-process \cite{Rauscher2013}.

As radioactive beam facilities in the future can address the remaining nuclear physics uncertainties in the p-process it will be interesting to see whether model discrimination becomes possible by comparing to solar and meteoritic p-process abundances. A promising step in this direction is recent work that demonstrated that the p-process synthesis of $^{94}$Mo occurs in a rather narrow temperature range. Once the nuclear physics is fully understood, the known natural abundance of this isotope can then be used to validate specific aspects of the conditions in the explosion. It would be interesting to systematically investigate the sensitivity of various p-isotopes to temperature and density conditions, and to specific features in the distribution of seed nuclei, and identify the ones with the most power towards model discrimination. Another focus should be the nuclear uncertainties affecting the production of the long lived radioactive p-isotopes $^{92}$Nb and $^{146}$Sm, which need to be minimized to use meteoritic data for validation. Future nuclear physics studies could then concentrate on the relevant reaction pathways that produce these key isotopes. 

\subsection{Nova models}

Nova explosions occur on the surface of white dwarfs that transfer matter onto their surface from a companion star in a stellar binary system. The nova explosion leads to a dramatic brightening of the star for many months \cite{Jose2007}. It is powered by proton capture reactions on stable and unstable neutron deficient isotopes. Validating nova models is important to understand the evolution of accreting white dwarfs in general, especially as such systems (not necessarily the ones that exhibit novae though) may be a pathway to explain type Ia supernova explosions. Compared to other astrophysical sites, nova models are in a somewhat curious position in terms of model validation. Despite of the strong sensitivity on difficult to determine nuclear reactions on unstable nuclei, the nuclear physics of nova explosions is relatively well understood thanks to decades of measurements at stable and radioactive beam facilities \cite{Iliadis2002}. The main reason is that the unstable nuclei are relatively close to stability and can be produced with sufficient intensity, and that temperatures during the explosion are high, resulting in much larger and easier to measure reaction rates compared to other stellar environments. Nevertheless, some key reactions in the most extreme class of novae, explosions on the surface of more massive oxygen-neon-magnesium white dwarfs, remain to be determined, especially in the context of the synthesis of heavier elements such as sulfur. An example is the rate of proton capture on unstable $^{30}$P, which serves as a bottleneck for the synthesis of heavier elements and needs to be known to use observations of sulfur abundances for model validation. As the development of radioactive beams progresses with new techniques and facilities, these remaining measurements should become feasible soon. It is possible that there is a subset of rare nova events where accretion rates and white dwarf temperatures lead to the accumulation of significantly more matter, with more extreme explosions and significantly extended nucleosynthesis \cite{Glasner2009}. In such events, a much broader range of nuclear reactions would become important to understand in order to predict observables that could be used to identify such events. 

The sensitivity of the observed composition of nova ejecta to conditions during the explosion has been investigated in detail, paving the way for using these observations as, for example, temperature probes of the exploding layer \cite{Downen2013}. This delineates a clear pathway for how model validation can be performed. However, a major challenge in the context of nova model validations are limitations in observational data. The number of novae with accurately determined abundance patterns that cover a broad range of elements is small. One issue is the limited availability of space based observatories with UV spectroscopy capability. Furthermore, such composition measurements are difficult to interpret and have large uncertainties, because of the rapidly evolving and complex structured ejecta cloud. This could be circumvented by detecting the $\gamma$-radiation of radioactive isotopes produced in the explosion, which penetrates more easily the ejected material \cite{Hernanz2006}. There are a number of suitable isotopes such as $^{22}$Na, or, in the early stages of the explosion, $^{18}$F. However, current $\gamma$-ray observational capabilities in the MeV range are not sufficient to make meaningful measurements of novae at typical distances. The analysis of pre-solar grains that are embedded in meteorites and may originate from dust formed in the ejecta of nearby novae provides a tantalizing opportunity to perform more detailed composition measurements. However, the nova origin of a particular grain is difficult to establish - more precise nuclear physics may help guide this identification process through comparison with the expected isotopic abundance pattern. Nevertheless, one has to be careful that this does not become a circular argument, where nova models are used to identify nova grains, which are then used to validate the nova model. 

\subsection{Other Theoretically Predicted Nucleosynthesis Sites}
Theoretical models sometimes predict novel sites of nucleosynthesis. Whether these predicted processes, or their sites, exist at all  must then be determined empirically. Nuclear physics is needed to make accurate predictions of the characteristic patterns of elemental or isotopic abundances produced in these hypothetical sites that one can then search for in observations. An example are so called Thorne-{\.Z}ytkow objects - red giant stars with a neutron star core where a pulsed or interrupted rapid proton capture process may produce anomalous abundance patterns of Rb, Sr, Zr, and Mo \cite{Biehle1991}. While theoretical models indicate that stable Thorne-{\.Z}ytkow objects cannot form, the discovery of stars with the predicted characteristic signatures have been recently reported \cite{Levesque2014}. In addition, it may be that even a temporary formation of a Thorne-{\.Z}ytkow like object during a merger of a star and a neutron star may produce interesting abundance signatures. It would therefore be useful to take advantage of recent improvements in our understanding of the nuclear physics of the rp-process to improve predictions of the abundance signatures produced by Thorne-{\.Z}ytkow objects. Other examples for theoretically predicted nucleosynthesis sites include the neutrino-p process in proton rich neutrino driven winds emerging from the nascent neutron star in core collapse supernovae \cite{Frohlich2006} or the extreme nova explosions predicted to occur on particularly cold accreting white dwarf stars \cite{Glasner2009}.  Also in these cases, the search for the signatures of these events has to be guided by predictions based on reliable nuclear physics. 

\subsection{Chemical Evolution Models}
There has been tremendous progress in astronomical observations of the compositions of stars. As the number of stellar composition data continues to grow dramatically, astrophysical models of single nucleosynthesis sources are not sufficient anymore to link nuclear physics with these observations and to take advantage of the wealth of additional information on abundance variations and correlations. Each star's composition reflects the composition of the Galaxy at the time and location of its formation (for the subset of stars were the composition has not been modified at a later stage, for example through mass transfer in a binary system). This composition is determined not only by a multitude of nucleosynthesis processes, but also by how ejecta from nucleosynthesis sites are mixed over time, which in turn depends on the complete merger history of the various stellar components that formed our Galaxy. Models of individual nucleosynthesis events must therefore be integrated into a model of Galactic chemical evolution to enable comparison, in a statistical sense, to the complete set of observational data. There has been significant progress recently in developing chemical evolution models based on N-body simulations of Galaxy formation \cite{Mosconi2001,Gomez2014,Hirai2015}. 

Improved nuclear physics in connection with nucleosynthesis calculations for a wide range of stellar masses and metallicities (so called "yield grids") in connection with the new observational datasets offers the opportunity to validate these models, our understanding of chemical evolution, and come to a true understanding of the origin of the elements, one of the ultimate goals of this field. This significantly broadens the scope of nuclear astrophysics, which becomes a tool to address questions of Galaxy formation, star formation, and galactic flows of matter. Instead of a simple comparison of abundance patterns, this validation now also considers the frequency of the nucleosynthesis event in question, how it depends on stellar parameters, and how the ejected abundance patterns vary over the history of the Galaxy. Such validation will be key for identifying convincingly sites of nucleosynthesis such as the r-process site \cite{Argast2004,Hirai2015,Shen2015}. 

One may argue that the large uncertainties in chemical evolution models and the large number of unknown parameters, that are then convoluted with the considerable uncertainties in the stellar nucleosynthesis sources, make it impossible to draw meaningful conclusions. Recent work has demonstrated that this is not necessarily so. Rather, the large number of observable elements, each probing different types of environments and sites, allows to control some of the parameter degeneracy. For example, a recent analysis demonstrated that deficiencies in stellar yield grids, and therefore in stellar models, can be identified in chemical evolution models even when taking into account the uncertainties of the chemical evolution process \cite{Cote2015}. 

Instead of looking at the chemical evolution of an entire galaxy, it is more straight forward to investigate smaller, to some extent self-contained, sub systems such as Globular Clusters or Dwarf Galaxies. An example is recent work that took advantage of experimental progress in reducing the nuclear physics uncertainties in stellar reaction rates achieved at stable beam facilities to explain the anomalous abundance trends observed in the globular cluster NGC2419 \cite{Iliadis2016}. 

\section{Viewing the Invisible - Probing Stars with Nuclear Physics}

Closely related to model validation is the extraction of parameters of a stellar system using precision nuclear physics and observations. This works best for models that have been validated for cases where all parameters are known, and are then applied to unknown systems. In this approach, nuclear physics essentially leads to a dramatic enhancement of the insights that can be gained from observations. 

\subsection{Probing the First Stars}
One of the most exciting examples of this are attempts to observe, characterize, and understand the very first stars that formed from the chemically pristine material created by the Big Bang \cite{Frebel2015}. While advanced infra-red telescopes such as the planned James-Webb Space Telescope attempt to provide this information by pushing observations to the faintest and most distant (and red-shifted) objects, observations of the composition of nearby stars in our Galaxy in combination with nuclear physics open an alternative pathway. This approach has been termed "Near Field Cosmology" because of the constraints that can be obtained on the environment in the early universe. Indeed it has been proposed that a class of rare carbon enhanced metal poor (CEMP) stars without detectable levels of neutron capture elements (CEMP-no) may have formed from the debris of the first supernova explosions after the Big Bang. With accurate nuclear physics, the observed composition can be matched to supernova models, where parameters such as the initial mass of the exploding star and the explosion energy can be varied to extract some of the properties of these early stars. An example is SMSS J031300.36-670839.3  where in addition to carbon also magnesium and calcium have been detected \cite{Keller2014}. Based on the pattern of the detected chemical abundances it was suggested that a relatively weak explosion of a first star with an initial mass of 60 solar masses created the observed signature. However, the nuclear processes responsible for the calcium production should be investigated and nuclear uncertainties addressed. In addition to hot hydrogen burning it was also proposed that a mild rp-process may occur in such massive early stars, albeit with larger mass \cite{Takahashi2014}. 

\subsection{Probing the Sun}
A closer to home example for how nuclear physics can be used to peek inside a star are plans to determine the metal content in the interior of the Sun. This can be done by using precise nuclear rates for the reactions in the CNO cycle and observations of neutrinos emitted from these reactions deep in the core of the Sun. With the well known solar temperature and density profiles, the amount of CNO nuclei can then be determined. This is interesting because the composition inferred from absorption spectra from the solar photosphere using sophisticated 3D atmosphere models \cite{Asplund2009} is in disagreement with helioseismic data \cite{Basu2008}. Neutrino experiments such as BOREXINO are improving their backgrounds with the goal to achieve a first measurement of the flux of CNO neutrinos from the Sun \cite{Davini2016}. At the same time, work is underway to determine with sufficient precision the proton capture rates in the CNO cycle \cite{Wiescher2010}. The determination of such stellar reaction rates has been a major challenge for the field because of the very small rates. 
 So far, with very few exceptions, none of the nuclear reactions in stars have been measured at the relevant energies, rather experiments are performed at higher energies and theory is used to extrapolate. This challenge is now being addressed with new technical developments, including the installation of high intensity accelerators underground (see below).
 
 \subsection{Probing Neutron Stars}
Neutron stars are very special objects that truly straddle the boundary between nuclear physics and astrophysics \cite{Chamel2008}. As km sized blobs of nuclear matter, they squarely fall into the realm of nuclear physics, though as astrophysical objects in space they are studied by astronomers with astronomical techniques instead of accelerators and detectors. The determination of neutron star properties from observations therefore not only provides information on stellar physics, but on the fundamental properties of nuclear matter. The recent discovery of a neutron star with a mass of about two solar masses immediately ruled out large classes of possible equations of state for nuclear matter \cite{Demorest2010} because it requires that compressed nuclear matter can provide enough pressure to balance the gravity for such a massive object. 

Accreting neutron stars in stellar binary system accumulate matter that is transferred from a regular companion star. The mass transfer onto the neutron star creates a wealth of observables which can be linked through accurate nuclear physics data to neutron star and nuclear matter properties  \cite{Schatz2006a,Parikh2013}. As the mass transfer rate varies, and sometimes stops entirely, the response of the neutron star can be observed, making accreting neutron stars a unique laboratory to study dense states of matter. 

One key observable are type I x-ray bursts, thermonuclear explosions of the matter that accumulates on the neutron star surface that repeat within hours or days and are the most frequent thermonuclear explosions observed in astrophysics. X-ray burst light curves and the spectral evolution during X-ray bursts has been used to constrain masses and radii of the underlying neutron star but the large range of unknown system parameters such as the composition of the accreted matter are a problem \cite{Oezel2016,Steiner2013,Lo2013}. It turns out though that the 10-100 s long X-ray light curve is sensitive to these parameters. The light curves are powered by the $\alpha$p- and rapid proton capture processes (rp-process) \cite{Schatz1998,Schatz2001,Parikh2013}. Once the nuclear physics is sufficiently well understood, models of burst light curves can be compared to observations of a particular source to constrain accretion rate, accreted composition, peak luminosity (and distance), and  thermal conditions of the neutron star surface. With these constraints, neutron star properties can then be extracted, and other open questions can be addressed. With still higher nuclear precision it may even be possible to use the shape of the light curve directly to probe gravity on the neutron star surface \cite{Zamfir2012}. 

The observation of the elements produced in X-ray burst nuclear reactions would open another very powerful pathway to understand accreting neutron stars. In particular, the redshift of identified spectral features created by these elements would provide direct constraints on the neutron star compactness. So far, no clear elemental signatures have been extracted from X-ray spectra - an example for a recent tentative measurement is the report of a possible chromium feature observed in GRS 1741.9-2853 by the NUSTAR satellite \cite{NUSTAR2015}. On the other hand, theoretical studies indicate that mass ejection of burst ashes via stellar winds should be expected for a subset of particularly powerful bursts \cite{Weinberg2006}. What is ejected is not the final burst ashes, but the intermediate composition created at the beginning of the burst at the time where the convection zone has its maximum extent (unless subsequent bursts are mixing up material from previous bursts). While frequency and ejected mass of such bursts are probably too low for a significant contribution to the elements found in the solar system, the resulting spectral features are predicted to be strong enough to be detectable with current instruments. It will be important to continue the search for spectral features in X-ray bursts, and theoretical work on predicting the expected signatures. 

Great strides have been made by the nuclear physics community to determine the nuclear reactions on very neutron deficient radioactive isotopes in the rp- and $\alpha$p-processes. Masses and decay properties are key quantities that have mostly been determined by measurements at radioactive beam facilities over the last decades \cite{Schatz2006a,Parikh2013}. However, the remaining challenge for the future are uncertainties in proton and $\alpha$-capture rates.
 
Another important observable for accreting neutron stars is the long term cooling of the crust that was heated by nuclear reactions during accretion. This cooling can be observed over many years in transient accretors where the accretion shuts off for years or decades \cite{Turlione2015}. A dozen or so systems have been observed so far.  The cooling behavior probes the physics of dense matter in the neutron star crust and may be sensitive to various phase transitions, including the presence of superfluid neutrons \cite{Brown2009} and nuclear pasta \cite{Horowitz2015}. However the quantitative comparison of  model curves with observations to extract this information requires a good understanding of the electron capture, $\beta$-decay, and fusion reactions of very neutron rich rare isotopes that heat and cool the neutron star crust during the accretion phase \cite{Haensel2008,Schatz2014}. All the critical nuclei become now accessible at the new generation of radioactive beam facilities and experiments geared towards addressing the nuclear physics questions related to neutron star crusts are an important new science motivation. Which reactions are the important ones depends also sensitively on the composition created by the X-ray bursts during the accretion phase. These burst ashes replace the original neutron star crust and set its composition. The nuclear physics of X-ray bursts is therefore another important element in the context of neutron star crust studies. A possible strategy could be to use precise $\alpha$p- and rp-process nuclear physics to fit burst models to the observed burst light curves of a particular transiently accreting system to predict that system's crust composition. This information can then be combined with experimental and theoretical nuclear data on extremely neutron rich nuclei to calculate crust cooling models, that can then be compared to observations during the accretion-off state to probe the dense matter inside the neutron star crust. 

\section{Nuclear Physics Enables Nucleosynthesis Predictions}
Beyond model validation and the probing of stellar environments, nuclear physics is key to predict individual nucleosynthesis contributions to elements and isotopes for which its not possible to disentangle processes observationally. In this case, models that have been validated otherwise (maybe with different processes or using different elemental abundance ranges) have to be outfitted with precise nuclear physics to predict the possible abundance pattern or patterns that a particular process may contribute. These predictions can then be used to disentangle the contributions to observed abundance patterns using chemical evolution models that also take into account the frequency, spatial distribution, and amount of ejected material of the various different stellar nucleosynthesis sources, as well as the time and space variability of these quantities. 

An important example are the "neutron capture" elements from germanium to about palladium. Observations of abundance patterns in metal poor stars clearly indicate that multiple processes contribute in this element range, and that the observed patterns are to some extent random mixtures of these processes \cite{Montes2007,Hansen2012,Hansen2014}. The processes that potentially contribute in this element range may include the classical weak s-, and main s-processes  \cite{Kaeppeler2011}, a main r-process, a weak r-process, an explosive charged particle process \cite{Hallmann2013}, the neutrino-p process in core collapse supernovae \cite{Frohlich2006}, or the Light Element Primary Process (LEPP) invoked to explain deficiencies in standard s-process models \cite{Travaglio2003}. These processes span stable, neutron rich, and neutron deficient isotopes and a wide range of nuclear reaction types. The relevant reaction paths are not as far from stability as for the r-process producing heavier elements, and fission processes from the main r-process are not expected to contribute strongly. This makes the nuclear physics of the synthesis of elements in the germanium to palladium range an important target for current and future stable, neutron beam, and radioactive beam facilities where it should be possible to obtain complete nuclear data sets that will enable an understanding of the origin of these elements. 

Another example is the prediction of the nucleosynthesis contribution to individual isotopes. In most cases, stellar observations are limited to elemental abundances and can therefore not be used to disentangle contributions to individual isotopes. An example are the s- and r-processes, where the main source of information on isotopic abundance patterns comes from the solar system composition, where both processes are mixed. While so called s- or r-only isotopes (isotopes that can only be reached by one of the two processes) and observed elemental abundances can provide guidance and serve as model normalization points, isotopic abundances inbetween these data points have to be interpolated using nucleosynthesis models and accurate nuclear physics. In the past, high precision information on s-process neutron capture rates has been key in predicting the s-process contributions to isotopes beyond iron, enabling the extraction of the predicted isotopic r-process abundances \cite{Arlandini1999}. In the future it may become possible with accurate r-process nuclear physics from advanced radioactive beam facilities to do the same for the r-process. This would allow to search for additional nucleosynthesis contributions beyond the s- and r-processes. 

\section{Maximizing the impact of nuclear physics work}

\subsection{Sensitivity studies and uncertainties}

Studies of the sensitivity of astrophysical observables to nuclear physics provide the intellectual connection between nuclear physics and astrophysics. They provide a deeper understanding on how nuclear physics and astrophysics are connected. They are also essential in assessing the impact of nuclear uncertainties on model validation, extraction of stellar parameters, and predictions of nucleosynthesis and serve therefore as the primary guide to a prioritized and high impact nuclear physics effort aimed at reducing these uncertainties. Often important nuclear physics data are well within reach of nuclear experiments or theory, but are not pursued because there is no strong nuclear physics motivation, and their astrophysical relevance is either not know, or not clear enough to convince program advisory committees at oversubscribed experimental facilities. Sensitivity studies directly address this issue and are therefore key to drive progress in nuclear astrophysics. 

A typical approach is to repeat astrophysical model calculations numerous times while varying the nuclear physics input. The computationally least expensive method is to vary individual nuclear reaction rates one by one. This is often the only feasible method, and it has the advantage that it clearly identifies the particular nuclear reaction that causes a change in the observables. It also provides a clearly defined result - the numerical sensitivity, essentially the derivative of the observable with respect to a particular reaction rate. The approach does however miss correlations between sensitivities - for example, a change of a reaction rate may lead to a change in the reaction path, and therefore change the importance of other reactions. An alternative is to vary pairs of reactions, or, the computationally most expensive approach, a Monte Carlo variation of all reactions. Monte Carlo studies have the advantage that they take into account all correlations (except for correlations in the input uncertainties) and that they can be used to properly propagate errors quantitatively to the predicted observables. However, they are computationally expensive, and require reasonable quantitative knowledge of uncertainties - unlike for single rate variations where the size of the change is a purely numerical tool to quantify sensitivity. It is also more difficult in Monte Carlo studies to trace changes in the observable back to the reaction rate uncertainties that caused them. This is usually done by searching for correlations between a given rate value and an observable. However statistical scatter can make this difficult to extract, especially when a large number of reactions contribute. 

Monte Carlo studies, as described, also have the drawback that they assume reaction rate uncertainties are uncorrelated. This is a reasonable assumption for processes where nuclear physics input is mostly experimental, such as the s-process, novae, or stellar processes. However, for processes where the uncertainties are dominated by uncertain theoretical reaction rates, such as  the i-, r-, rp-, $\nu p$, or p-processes, uncertainties can be highly correlated. For example, many ($\alpha$,p)-reactions in the $\alpha p$ process have been calculated using the statistical Hauser-Feshbach theory. It has been suggested that because of a too low level density, uncertainties in $\alpha$-potentials, and $\alpha$-clustering effects, ($\alpha$,p) reaction rates may be systematically overpredicted by up to two orders of magnitude. Similarly, shell effects in the r-process can lead to systematic changes of neutron capture rates, masses, and $\beta$-decay rates which may lead to much larger changes in the produced abundances than an independent Monte Carlo variation may indicate. A possible solution is to randomly vary not reaction rates, but the input into the theoretical calculations (see for example \cite{Bertolli2013}). The drawback here is that one is limited to the correlations imposed by a particular theory, which may not be applicable in the first place. Correlations due to deficiencies in the theory used are still missing in such an analysis. 

Comprehensive sensitivity studies have been carried out for a wide range of astrophysical sites and processes. Systematic r-process sensitivity studies over a broad range of r-process sites and nuclear physics quantities have been particularly influential, and have provided strong support for the scientific motivation of FRIB \cite{Mumpower2016}. It has also recently become possible to run a full Monte Carlo sensitivity analysis of a complete stellar evolution model for low mass stars from stellar birth to the formation of the final carbon-oxygen core \cite{Fields2016}. Other examples for large scale sensitivity studies include studies for the p-process \cite{Rapp2006}, novae \cite{Iliadis2002}, X-ray bursts \cite{Parikh2008,Parikh2009,Cyburt2016}, the s-process \cite{Koloczek2016}, the i-process \cite{Bertolli2013}, the Sun (see \cite{Adelberger2011} for a discussion of the relevant reactions), explosive nucleosynthesis in general \cite{Woosley1986}, weak interactions in core collapse supernovae \cite{Sullivan2016}, and the Big Bang \cite{Cyburt2015}. Many other sensitivity studies have been carried out for specific reaction rates, or for more specific observables such as the synthesis of particular isotopes. 

It is important to understand the limitations of sensitivity studies. A sensitivity study is carried out with a particular astrophysical model code, and for a specific set of astrophysical parameters. The nuclear physics identified as "important" is therefore specific to the particular model investigated. In nature, the conditions in astrophysical sites can vary widely - for example, X-ray burst properties vary widely from system to system, depending on the accretion rate, accreted composition, and neutron star gravity and therefore nuclear reaction sequences are expected to be different. Sensitivity studies therefore have to be carried out for a grid of parameters that span the realistic range of the particular astrophysical site. In addition, models from different groups for the same astrophysical scenario differ in their choices for approximating complex stellar processes and in the level of sophistication in the description of various model aspects. Sensitivity studies carried out with models from different groups can therefore differ in their results. This is sometimes misunderstood as a problem that prevents meaningful nuclear physics work. This is however not so. There is no reaction that is "important for X-ray bursts". Rather, the nuclear physics identified in a sensitivity study is the  data set that is needed to validate one particular model of one particular group. Nuclear scientists provide the data needed by various models, and comparison with observation can then help guide improvements of the astrophysics, falsification of some models, and lead to new astrophysical insights. The important nuclear physics may change as a result of this process and sensitivity studies may have to be repeated. 

Another important issue related to sensitivity studies is the choice of observable, or the figure of merit that quantifies the change in the astrophysical model caused by a change in nuclear data. The results of the study, and the prioritization of the nuclear physics uncertainties, depend strongly on this choice. For example, the most critical nuclear physics for an r-process study that aims at disentangling the origin of Y, Sr, and Zr may not even be included in an r-process sensitivity study that uses the number of abundances affected over the entire r-process range as a figure of merit. This issue deserves more attention. In the future it would be important in sensitivity studies to carefully identify as much as possible the potential questions a nucleosynthesis model seeks to address, and provide multiple tailored priority rankings. 
	
\subsection{Databases and Evaluations} 

Dedicated astrophysical databases are critical for progress in nuclear astrophysics. Just like in nuclear science, having all relevant nuclear data available in a consistently evaluated database is key to ensure the best possible data are used, and model calculations are carried out with well defined and reproducible data sets. In the case of nuclear astrophysics, this is even more important because of the interdisciplinary nature of the field. Astrophysicists do not have the expertise to evaluate nuclear data for thousands of reactions, so if a nuclear physics result is not made available in a data base chances are it will never be used for a nuclear astrophysics problem. In addition, published nuclear data are seldom in a form that is usable in an astrophysical model. Usually data from a range of experiments have to be combined, and complemented with theoretical calculations that extend the data to temperature regimes not covered by experiment. In addition, corrections need to be applied for the high temperatures and densities in astrophysical environments. Nuclear astrophysics therefore requires a very different type of database than nuclear physics, and in particular requires a much more extensive effort to compile and transform the raw data into a usable form. 

Currently the field is served by a range of database efforts that cover different types of data and have different priorities and approaches. There are currently three databases that compile complete sets of reaction rates for astrophysical applications, including experimentally based rates and theoretically predicted rates: BRUSLIB\footnote{\href{http://www.astro.ulb.ac.be/bruslib/}{http://www.astro.ulb.ac.be/bruslib/}} maintained by the Bruxelles nuclear astrophysics group \cite{Xu2013}, JINA REACLIB\footnote{\href{https://groups.nscl.msu.edu/jina/reaclib/db/}{https://groups.nscl.msu.edu/jina/reaclib/db/}} maintained by the Joint Institute for Nuclear Astrophysics JINA-CEE \cite{Cyburt2010}, and the more recent STARLIB\footnote{\href{http://starlib.physics.unc.edu}{http://starlib.physics.unc.edu}} maintained by the University of North Carolina Chapel Hill nuclear astrophysics group \cite{Sallaska2013}. BRUSLIB offers a wide range of data in addition to reaction rates, such as level density, strength functions and ground state properties. JINA REACLIB provides continuously updated reaction rates in a format that uses temperature dependent fits, and offers version tracking and pre-defined or user defined library snapshots. STARLIB offers complete published recommended data sets in table format and is the only library that also includes estimates for the reaction rate uncertainties. None of these libraries integrate temperature and density dependent weak interaction rates, which would be an important development for the future. 

In addition there are more targeted evaluation and database efforts that focus on specific types of reactions. Often the results of these compilations are included in the comprehensive databases mentioned above. Detailed evaluations were performed by the NACRE collaboration for reactions on stable target nuclei in the $A=1-28$ mass range and published in 1999 \cite{NACRE1}. In 2013 an update was published that covers the $A=1-16$ mass range \cite{NACRE2}. The National Nuclear Data Center evaluated neutron capture rates in 2010 \cite{Pritychenko2012}. A more continuos compilation effort for neutron capture rates is performed by KADONIS \cite{KADONIS2014} \footnote{\href{http://www.kadonis.org}{http://www.kadonis.org}}. KADONIS also compiles experimental data for proton and $\alpha$ induced reactions in the germanium to bismuth elemental range for p-process studies. The JINA-CEE/NSCL  weak rate library\footnote{\href{https://groups.nscl.msu.edu/charge_exchange/weakrates.html}{https://groups.nscl.msu.edu/charge\_exchange/weakrates.html}} compiles weak interaction rates as functions of temperature and density based on theoretical and experimental data. nucastrodata.org at ORNL also contains various data sets of nuclear astrophysics data, and in addition offers a suite of tools to aid compilation and evaluation of astrophysical reaction rates. 

There are a number of important sources of nuclear data that are compiled for nuclear physics purposes but are of importance for astrophysical calculations. These include the Atomic Mass Evaluation (AME)\footnote{\href{https://www-nds.iaea.org/amdc/}{https://www-nds.iaea.org/amdc/}} for nuclear masses and nubase\footnote{\href{https://www-nds.iaea.org/amdc/}{https://www-nds.iaea.org/amdc/}}  and NNDC's Nuclear Wallet Cards\footnote{\href{http://www.nndc.bnl.gov/wallet/}{http://www.nndc.bnl.gov/wallet/}} for decay data. Branchings for $\beta$-delayed neutron emission are now being compiled and reevaluated by an effort within the framework of the IAEA and the results for the element range from $Z=1-28$ have recently been released \cite{Birch2015}. 

\subsection{Centers} 
\setcounter{footnote}{0}
Interdisciplinary centers such as the Joint Institute for Nuclear Astrophysics\footnote{\href{http://www.jinaweb.org}{http://www.jinaweb.org}}  in the US or the NAVI Virtual Nuclear Astrophysics Institute in Germany play an important role in nuclear astrophysics. They serve as centers for the field as a whole and forge communication, collaboration, and education across the field boundaries of nuclear physics and astrophysics. This is important for many reasons:  What is important to do in nuclear physics depends critically on the astrophysical frontiers questions one tries to answer using the nucleosynthesis models where the data are being used. This in turn depends on new observations, new astrophysical models, and may change quickly with new observational discoveries. Similarly, understanding the possibilities and issues in nuclear physics and nucleosynthesis in general can change priorities in astrophysics theory and observations. A joint, interdisciplinary approach also increases the focus on work that creates the connection between nuclear physics and astrophysics, such as sensitivity studies or even just the implementation of nuclear physics in astrophysical models, work that otherwise would be at the fringes of each field and often gets neglected. 

\section{Developments in Experimental Nuclear Astrophysics}

Nuclear physics work plays a key role in nuclear astrophysics. It enables model validation, the prediction of nucleosynthesis signatures and contributions, and greatly expands the information that can be extracted from astrophysical observations, including properties and parameters of stellar sites that would otherwise not be accessible. The two long standing nuclear physics challenges - the determination of the very low nuclear reaction rates in stars, and the determination of properties and reactions of very neutron deficient or very neutron rich unstable nuclei, are now being addressed with new technical developments. For stellar reaction rates, new or upgraded stable beam accelerators that focus on nuclear astrophysics are coming on line for example above ground St. Ana at  Notre Dame, LENA at UNC Chapel Hill, and FRANZ in Frankfurt Germany, and, to increase sensitivity by shielding experiments from cosmic ray background, deep underground at CASPAR in Sanford, USA \cite{Robertson2016}, or LUNA in Gran Sasso, Italy \cite{Formicola2014}. New techniques are being developed to further increase sensitivity, including the use of recoil separators (St. George at Notre Dame \cite{Couder2008}, ERNA at Naples \cite{Gialanella2012}), new detection systems for coincident summing (SUN at MSU \cite{Quinn2014} and new systems at UNC \cite{Buckner2015}), optical TPCs \cite{Gai2012}, bubble chambers \cite{DiGiovine2015} or indirect reaction techniques \cite{Tribble2014} such as the Trojan Horse method or Coulomb Breakup. At the same time, the field is addressing the production of radioactive beams needed for astrophysics studies with a new generation of radioactive beam facilities in the US, Canada, Europe, Japan, China, and Korea, for example the FRIB facility in the US. These new facilities not only dramatically increase the reach of produced isotopes, but also broaden the range of production mechanisms used to create low energy astrophysics beams for direct reaction rate measurements. Developments in this context include the new ReA3 facility at NSCL/FRIB in the US, which uses radioactive beams produced by fragmentation, stops the beam in a gas cell, and then reaccelerates the beam particles to provide a high quality beam at low astrophysical energies. This facility is now operational and a regular experimental program has started in Fall 2015. An active target TPC (AT-TPC) \cite{Mittig2015} and a gas jet target (JENSA) \cite{Chipps2014} are available for nuclear astrophysics reaction studies. The SECAR recoil separator is currently under construction to enable direct astrophysical reaction rate measurements in conjunction with JENSA \cite{Berg2016}. Another approach is planned at FAIR in Germany, where fragmentation beams will be decelerated in a storage ring to perform astrophysical reaction rate measurements inside the ring \cite{Woods2015}. These new approaches complement the traditional ISOL technique with a post accelerator, for example at TRIUMF and ISOLDE. At TRIUMF, the DRAGON recoil separator facility will continue direct astrophysical reaction measurements as more radioactive beams are developed \cite{Hutcheon2008}. AT ISOLDE, the TSR storage ring is currently under construction to enable such measurements \cite{Grieser2012}. 

\section{Summary} With the significant developments in theoretical and experimental nuclear physics, the continued progress in astronomical observations and multi-dimensional computer simulations of stellar environments, and the strong connections between nuclear physics and astrophysics fostered by centers like the Joint Institute for Nuclear Astrophysics the field is in an excellent position for major progress in the coming decade. 

\section{Acknowledgements}
This material is
based upon work supported by the National Science Foundation under
Grant Numbers PHY-02-016783, PHY-08-22648, and PHY-1430152 (JINA
Center for the Evolution of the Elements).

\section*{References}
\bibliographystyle{iopart-num}
\bibliography{hsref_v4}
%\begin{thebibliography}{10}
%\bibitem{ref1} J.~Doe, Article name, \textit{Phys. Rev. Lett.}
%\end{thebibliography}

\end{document}